\documentstyle[12pt,psfig]{article}
\newcommand{\beq}[1]{\begin{equation}\label{#1}}
\newcommand{\eeq}{\end{equation}}
\newcommand{\beqar}[1]{\begin{eqnarray}\label{#1}}
\newcommand{\eeqar}{\end{eqnarray}}
\newcommand{\lash}[1]{\not\! #1 \,}
\begin{document}
\vspace*{-2cm}
\hfill UFTP preprint 380/1995
\vspace{4cm}
\begin{center}
{\bf \large  QCD Sum Rule Calculation of Twist-4 Corrections to
Bjorken and Ellis-Jaffe Sum Rules}
\vspace{1cm}

E.~Stein$^1$, P.~G\'ornicki$^2$, L.~Mankiewicz$^{3}$, and A.~Sch\"afer$^1$
\vspace{1cm}

$^1$Institut f\"ur Theoretische Physik, J.~W.~Goethe
Universit\"at Frankfurt,
\\
Postfach~11~19~32, W-60054~Frankfurt am Main, Germany
\\
$^2$Institute of Physics, Polish Academy of Sciences,
Al. Lotnikow 32/46, PL-02-668 Warsaw, Poland
\\
$^3$N. Copernicus Astronomical Center, Bartycka 18, PL--00--716 Warsaw,
Poland
\end{center}

\vspace{2cm}
\noindent{\bf Abstract:}
We calculate the twist-4 corrections to the integral of $g_1(x,Q^2)$ in
the framework of QCD sum rules using an interpolating nucleon field
which contains explicitly a gluonic degree of freedom. This
information can be used together with previous calculations of the
twist-3 contribution to the second moment of $g_2(x)$ to estimate the
higher-twist corrections to the Ellis-Jaffe and Bjorken sum rules.
We get $f^{(2)}(proton) = -0.037 \pm 0.006$ and
$f^{(2)}(neutron) = -0.013 \pm 0.006$.
Numerically our results roughly agree with those obtained by
Balitsky, Braun and Kolesnichenko based on a sum rule
for a simpler current.
Our calculations are far more stable as tested within the
sum rule approach but are more sensitive to less well known condensates.
\eject
\newpage
Spin effects in strong interaction high energy processes are one of the best
tools to study  QCD both in the perturbative and non perturbative regime.
It has become clear that the reliable determination of the
$Q^2$ dependence, both due to radiative corrections and due to higher twist
contributions is a central task of QCD theory.
The $Q^2$ dependence of spin variables is in general more benign than
for unpolarized quantities (e.~g.~the anomalous dimension for the
Bjorken sum rule vanishes), allowing to extract very interesting
information from data taken at $Q^2$ as low as 1 $\rm GeV^2$.
While the EMC \cite{EMC}
and SMC \cite{SMC} experiments have still comparatively large
$Q^2$ , SLAC \cite{SLAC} data are taken
down at a rather low mean $Q^2$ of about $2 \rm \; GeV^2$.
The proton data indicated a disagreement with the polarized-proton sum rule,
the Ellis-Jaffe sum rule \cite{EJ}, leading to a lot of excitement in
the high-energy physics community.
Since experiments now not only provide data for the
spin-dependent structure
function of the proton $g_1^p(x)$ but also of the neutron
$g_1^n(x)$ one of the most solid predictions of QCD, the
Bjorken sum rule \cite{Bj}, is tested experimentally.
However, for low $Q^2$ this sum rule (strictly valid in the asymptotic
Bjorken limit) receives corrections. The most familiar are those from
perturbative QCD, calculated for the leading twist term up to order
$\alpha_S^3$ \cite{larin} and estimated to order $\alpha_S^4$ with an
rough estimate of order $\alpha_S^5$ \cite{kataev}.
Higher-twist corrections are given in terms of complicated
hadronic matrix-elements and are suppressed by powers of $Q^2$.
Since experiments still are not in a save region of asymptotically
large $Q^2$ all this corrections have to be examined carefully to
give a complete picture of the $Q^2$ dependence of the Bjorken
sum rule \cite{Kar}.

The first moment of $g_1(x,Q^2)$ at fixed $Q^2$
is given by \cite{bruno}
\beq{eins}
\int_0^1 dx \; g_1 (x,Q^2)
=
\frac{1}{2} a^{(0)} + \frac{m_N^2}{9 Q^2}
\left( a^{(2)} + 4 d^{(2)} + 4 f^{(2)} \right) + O ( \frac{m_N^4}{Q^4} )
\; .
\eeq
In the above formula we have not included higher order corrections to
the coefficient functions which can be written as power series in
$\alpha_S(Q^2)$ \footnote{The radiative corrections to the
leading-twist singlet part were calculated to order $\alpha_S^2$
\cite{Larin94}
and estimated to order $\alpha_S^3$ \cite{Kataev94}.}
In the present paper we focus on the correction proportional to
$f^{(2)}$ which is defined by the matrix element of the twist-4
operator
\beq{Osigma}
O_\sigma(0) = \bar{q}(0) g {\tilde G}_{\sigma \beta} (0) \gamma^\beta q (0)
\eeq
in the nucleon state $|pS \rangle$ of momentum $p$ and spin $S$,
$p^2 = - S^2=m_N^2$, $p \cdot S = 0$,
\beq{f2}
\langle pS| O_\sigma(0) |pS \rangle = 2 m_N^2 f^{(2)} S_\sigma \, .
\eeq
Here and in the following we temporarily neglect the normalization-point
dependence of the operator $O_\sigma$.

The reduced matrix elements $d^{(2)}$ and $f^{(2)}$ can be expressed
through the second moments of the polarized nucleon structure
functions $g_1(x)$ and $g_2(x)$ \cite{Jaffe1}:
\beqar{d2}
\int dx \; x^2 g_2(x) &=& \frac{1}{3} \left( d^{(2)} - a^{(2)} \right)
\nonumber \\
\int dx \; x^2 g_1(x) &=& \frac{1}{2} \; a^{(2)}
\eeqar
While $a^{(2)}$ may be  taken directly from experiments, the matrix
elements $f^{(2)}$ and $d^{(2)}$ were first estimated by Balitsky,
Braun and Kolesnichenko \cite{BBK} using the QCD sum rules techinque.
In our previous work \cite{SGMS} we presented an independent sum rule
calculation of $d^{(2)}$ which essentially confirmed the values obtained
by BBK.
The calculation of
$f^{(2)}$ presented in this paper completes therefore the program of
estimating the leading power corrections to the sum rules for $\int
g_1(x,Q^2) dx$.

Before we are going to dwell on the details of the calculation we would
like to stress that the leading higher-twist matrix elements
describe fundamental properties of the nucleon.
The twist-4 operator eq.~(\ref{Osigma}) is a measure
for the contribution of the collective gluonic field to the spin of
the nucleon.
Writing the dual field strength tensor in its components we get
\beq{Ocomp}
\langle pS| - B^\sigma_A j_A^0 + (\vec{j}_A \times \vec{E}_A)^\sigma
|pS \rangle = 2 m_N^2 f^{(2)} S^\sigma
\eeq
where the quark-current is denoted as
$j_A^\mu = - g \bar q \gamma^\mu t^A q$ and $B_A^\sigma$ and
$E_A^\sigma$ are the colour magnetic and colour electric fields.
In the rest system of the nucleon an analogous relation holds for
the twist-3 operator which determines $d^{(2)}$
\beq{O3comp}
\langle pS| 2 B^\sigma_A j_A^0 + (\vec{j}_A \times \vec{E}_A)^\sigma
|pS \rangle = 8 m_N^2 d^{(2)} S^\sigma \; .
\eeq
Knowledge of $d^{(2)}$ and $f^{(2)}$ then allows to estimate
magnetic and  electric field contributions to the spin separately.

In the usual approach nucleon matrix elements of local operators can
be extracted from a three-point correlation function
\beq{dreipunkt}
\Pi_\Gamma(p) = i^2 \int d^4x e^{ipx} \int d^4y
\langle 0|T\left\{ \eta(x) O_\Gamma(y) \overline{\eta}(0)\right\}| 0 \rangle
\eeq
which involves an interpolating current $\eta(x)$ with a certain
overlap $\lambda$ between the state created from the vacuum by
$\eta(x)$ and the nucleon state
\beq{overlap}
\langle 0| \eta(x))|pS \rangle = \lambda u(p,S) \exp{(ipx)} \, .
\eeq
The overlap integral can be determined from an additional
two-point correlation function
\beq{zweipunkt}
\Pi(p) = i \int d^4x e^{ipx}
\langle 0|T\left\{ \eta(x) \overline{\eta}(0)\right\}| 0 \rangle \, .
\eeq
In practical application it is often advantageous to consider the
ratio of three- and two-point correlation functions such that the
$\lambda$-dependence cancels out.

For QCD sum rule calculations of nucleon properties the standard
choice for $\eta(x)$ has been for a long time the three-quark current
introduced by Ioffe \cite{Ioffe}
\beq{ioffec}
\eta_I(x) = \left[u^a(x) C \gamma_\mu u^b(x)\right]
                  \gamma_5 \gamma^\mu d^c(x) \varepsilon^{abc}\, ,
\eeq
which was used in the calculation of $f^{(2)}$ and $d^{(2)}$ by
Balitsky, Braun and Kolesnichenko (BBK) \cite{BBK}.  As explained in
\cite{BGMS,SGMS} for the investigation of operators which, like
$O_\sigma$ eq.~(\ref{Osigma}), contain explicitly gluonic degrees of
freedom it is very useful to match these by an interpolating nucleon
current that contains gluonic degrees of freedom as well.
For such a proton current we chose
\beq{qqqG}
\eta_G(x) =
 \frac{2}{3} \left(\eta^{\rm old}_G(x) - \eta^{\rm ex}_G(x)\right) \; ,
\eeq
where
\beq{qqqGold}
\eta^{\rm old}_G(x) =\varepsilon^{abc} \left(u^a(x) C \gamma_\mu
u^b(x)\right)
\gamma_5 \gamma^\mu \sigma_{\alpha \beta} \left[G^{\alpha \beta}(x) d(x)
\right]^c \; ,
\eeq
and
\beq{qqqGex}
\eta^{\rm ex}_G(x) = \varepsilon^{abc} \left(u^a(x) C \gamma_\mu
d^b(x)\right)
\gamma_5 \gamma^\mu \sigma_{\alpha \beta} \left[G^{\alpha \beta}(x) u(x)
\right]^c \; .
\eeq
This current was first studied in \cite{BGMS} and tested in the
calculation of the nucleon gluonic form factor and the total momentum
fraction carried by gluons.
Next the twist-3 correction $d^{(2)}$ occurring in the
expansion of $\int g_1(x) dx$ was predicted \cite{SGMS}. Note that the
current (\ref{qqqG}) may be regarded as the leading
expansion term of a non-local version of the
classical three-quark current (\ref{ioffec}). Working with non-local
sources has become also popular in lattice-gauge theories where
stability can be increased by taking a number of derivatives of
the quark-fields.

Using lattice-gauge theory the three- and two-point correlators
(\ref{dreipunkt}) and (\ref{zweipunkt}) can be calculated
directly from first principles. Such a project is pursued
by the J\"ulich-group \cite{Horsley}. The basic idea of the
QCD sum rule technique on the other hand is to employ the duality
between the hadronic and partonic representation of a correlation function
and to extract the quantity of interest by demanding that both
descriptions match each other at some intermediate scale. If $-p^2$ is
sufficiently large the main contribution comes from small
$x$-distances of order of $x^2 \sim 1/(-p^2)$. The consideration of
contributions from different t-channel distances is more
involved. When $y^2 \le 1/(-p^2)$ the standard machinery of the
short-distance expansion is applicable resulting in the known
expansion in terms of quark and gluon condensates.  However, the
contribution from large $y^2$  has to be accounted for separately.

The solution to this problem was first formulated by Balitsky
\cite{Bal1}. The Operator Product Expansion (OPE) of a three-point
correlation function has a twofold structure. Terms of the first type
come from the region $x^2 \sim y^2 \sim 1/(-p^2)$ and are proportional
to vacuum expectation values (VEV) of local gauge-invariant operators
multiplied by coefficient functions depending on $p^2$. In the
following we shall refer to these terms as to local power corrections
(LPC).

Terms of the second type called bilocal power corrections (BPC)
originate from distances $y^2 \gg x^2 \sim 1/(-p^2)$. To treat such
contributions properly one should expand the time-ordered product of
nucleon interpolating currents
\beq{ope1}
T(\eta_I(x){\bar \eta}_G(0)) = \sum_n C_n^{BL}(x) {\tilde O}_n(0) ,
\eeq
in a series of local, gauge-invariant operators ${\tilde O}_n(0)$ of
increasing dimension $n$. When this expansion is inserted back in
(\ref{dreipunkt}) it results, together with standard LPC, in the
following general form of the OPE of the three-point correlator
(\ref{dreipunkt})
\beq{ope}
\Pi_\Gamma(p) =   \sum_n c_{\Gamma,n}^{\rm L}(p) <O^{\rm L}_n>
                  + \sum_n c^{\rm BL}_n(p) \Pi_{\Gamma,n}^{\rm BL}(0)
\; .
\eeq
The bilocal power corrections are determined by the long-distance,
non-perturbative contributions to correlation functions at zero
momentum \cite{Bal2}:
\beq{twopoint}
\Pi_{\Gamma,n}^{\rm BL}(q) = i
\int d^4y e^{iqy} \langle 0|T\left\{O_\Gamma(y)
\tilde{O}_n(0)\right\}|0\rangle \, ,
\eeq
with $q=0$. Special care has to be taken to properly eliminate the
perturbative short-distance singularities from these correlators which
are already included in LPC coefficients $c_{\Gamma,n}^{\rm L}(p)$.
We stress again that the essence of the expansion (\ref{ope}) is the
separation of different scales: vacuum expectation values of local
operators $\langle O^L\rangle$ and correlators $\Pi^{BL}$ describe
long distance effects while the coefficients $c^L$ and $c^{BL}$
receive contribution only from highly virtual quark and gluon fields
which propagate for small distances. The OPE of the correlation
function $\Pi_\Gamma$ must be written as the sum of both LPC's and
BPC's, as above. We note that in general only this sum has a physical
meaning and is independent of the regularisation scheme. Contrary to
the case of our previous calculation of the twist-3 correction \cite{SGMS},
the BPC's do play a crucial role in the analysis of the twist-4 matrix
element.

The right hand side of (\ref{dreipunkt}) can be decomposed into
different Lorentz structures according to
\beqar{twist4}
&&
i\int d^4x e^{ipx} i \int d^4y \langle 0|T \left\{
\eta_I(x) \bar{q}(y) g\tilde{G}_{\sigma\mu}(y) \gamma^\mu q(y)
\overline{\eta}_G(0)
\right\}|0\rangle \nonumber \\
&&= p_\sigma {\lash p} \gamma_5 W^{\cal A}(p^2) + \gamma_\sigma \gamma_5
W^{\cal B} (p^2) + \ldots \, ,
\eeqar
where the ellipses stand for other terms which can be eliminated
taking the trace of (\ref{twist4}) with an appropriate
projector. The invariant functions $W^i$, $i={\cal A}, {\cal B}$
can be represented as
spectral integrals \cite{Bal2}
\beqar{spec}
W^i(p^2) &&= \int ds \frac{1}{s-p^2}\left[a^i \delta^\prime(s-m_N^2) +
b^i \delta(s-m_N^2) + \Theta(s-s_0) \rho^c(s)\right]
\nonumber \\
 &&+ \; {\rm subtractions} \, .
\eeqar
We have accepted the conventional ``resonance plus continuum'' model
of the spectral density with the continuum density $\rho^c(s)$ dual to
all graphs with non-vanishing imaginary part at $-p^2 \to \infty$. The
constant $a^i$ which stands in front of the double-pole term is
proportional to $f^{(2)}$ while the single-pole term is determined by
nucleon-to-continuum transitions and has to be eliminated from the
final answer. In principle the information about the magnitude of
$f^{(2)}$ can be extracted either from $W^{\cal A}$ or from $W^{\cal
B}$.  In practical calculations it is advantageous to consider
$W^{\cal A}$ because of its lower dimensionality and this structure
was chosen for the analysis presented in \cite{BBK}. We realize that
the QCD sum rule approach involves a number of approximations the
accuracy of which is sometimes difficult to assess a'priori. To get a
better feeling of the intrinsic uncertainties of the whole method we
have decided to analyze $W^{\cal B}$ which leads to a sum rule which
is more sensitive to the higher mass region in the spectral
representation (\ref{spec}).  The constant coefficient
if front of the double-pole nucleon contribution to
$W^{\cal B}$ can be found to be equal to
\beq{NC}
a^{\cal B} = - 2 f^{(2)} \lambda_I \lambda_G m_N^6 \; .
\eeq
Here, $\lambda_I$ and $\lambda_G$ are the overlap of the Ioffe current
(\ref{ioffec}) respectively the quark - gluon current (\ref{qqqG}) with
the nucleon, while $m_N$ denotes the nucleon mass.

The invariant function $W^{\cal B}$ can be easily projected out of the
three-point correlator (\ref{dreipunkt}) by taking the trace with
$\frac{1}{4} \gamma_5 {\lash S}$ and choosing momentum $p_\mu$ such that
$p\cdot S = 0$. The expansion of $W^{\cal B}$ in LPC is standard and we are
not going to dwell on the details here. The net result can be written
as the following formula:
\beqar{twist4ope}
&& 4 W^{\cal B}_{\rm LPC}(p^2)
= \nonumber \\
&&
A \frac{\alpha_S}{\pi^5} (-p^2)^4 \log(-p^2/\mu^2)
+ B \frac{\alpha_S}{\pi} <\bar{q} q>^2 (-p^2) \log(-p^2/\mu^2)
\nonumber \\
&&
+ C \frac{1}{\pi^4} <fg^3GGG> (-p^2) \log(-p^2/\mu^2)
\nonumber \\
&&
+ F \frac{\alpha_S}{\pi} m_0^2<\bar{q} q>^2
\nonumber \\
&&
+ G \frac{<\bar{q} q>^2 <g^2GG>}{-p^2}
+ H \frac{<\bar{q} q>^2 m_0^4}{-p^2}
\eeqar
where the numerical coefficients are given in Table \ref{tabelle1}.
Note that the coefficient in front of the gluon condensate $\langle
GG \rangle$ vanishes.
\begin{table}
\begin{center}
\begin{tabular}{||l|l|l|l|l|l|l||}
\hline \hline
  &  $A$  &  $B$ & $C$ & $D$ & $G$ & $H$ \\ \hline
d & $ -1/2592 $ & -4/3  & $0 $ & 88/81 & 5/81 &
1/9  \\ \hline
u & $ -1/3240 $ & -20/27   & $-23/2304$ & 1411/2592 &  1/81 &
-1/54 \\ \hline
\hline
\end{tabular}
\end{center}
\caption[]{\sf Numerical coefficients corresponding to LPC contributing to
the sum rule eq. \ref{twist4ope} .
The upper line gives the values for the twist-4 operator involving d quarks,
the lower line for the corresponding operator with u-quarks}
\label{tabelle1}
\end{table}

The calculation of BPC's is more involved and will be described next in
some details. Let us consider operators which may contribute two-point
correlators (\ref{twopoint}) of dim-6 and dim-10 to the sum rule. The
expansion of $T(\eta_I(x){\bar \eta}_G(0))$ in local operators leads
to the following series:
\newpage
\beqar{etaope}
&& T(\eta_I(x){\bar \eta}_G(0)) =
        \frac{20 x_\sigma}{3 \pi^4 x^8} \gamma_5 \lash{x}
	\left[i \left(\bar d \gamma_5 \gamma_\varrho gG^{\varrho\sigma}d \right)(0)
	      - \left(\bar d \gamma_\varrho g\tilde{G}^{\varrho\sigma}d \right)(0)
        \right]
\nonumber \\
&&+
\frac{16 x_\sigma}{3 \pi^4 x^8} \gamma_5 \lash{x}
        \left[i \left(\bar u \gamma_5 \gamma_\varrho gG^{\varrho\sigma}
u\right)(0)
              - \left(\bar u \gamma_\varrho g\tilde{G}^{\varrho\sigma}
u\right)(0)
        \right]
\nonumber \\
&&+
 m_0^2 <\bar{q} q> \frac{x_\alpha x_\lambda}{6 \pi^2 x^4}
  \gamma_\varphi \gamma_5
  \left[ 2 \left(\bar u \gamma_5 \sigma^{\varphi\lambda} D^\alpha u\right)(0)
    - \frac{1}{3}\left(\bar d \gamma_5 \sigma^{\varphi\lambda} D^\alpha
d\right)(0)
\right.
\nonumber \\
&&-
\left.
i g^{\varphi\lambda} \left(\bar d \gamma_5 D^\alpha d \right)(0)
  \right]
\nonumber \\
&& -  \alpha_S <\bar{q} q>^2
\frac{8}{27 \pi x^4} \left(5 x^\varphi x_\alpha - 9 g^\varphi _\alpha
x^2\right)
\gamma_\varphi \gamma_5
\left(\bar u \gamma^\alpha \gamma_5 u\right)(0)
\nonumber \\
&& - \alpha_S <\bar{q} q>^2
\frac{32 x^\varphi x_\alpha}{27 \pi x^4}
\gamma_\varphi \gamma_5
\left(\bar u \gamma^\alpha \gamma_5 u\right)(0)
\nonumber \\
&&-
<\bar{q} q> \frac{4 i }{9 \pi^2} \frac{x_\lambda x_\alpha}{x^4}
\gamma_\varphi \gamma_5
\left[\left(\bar u g\tilde{G}^{\lambda\varphi} D^\alpha u\right)(0) +
     2\left(\bar d g\tilde{G}^{\lambda\varphi} D^\alpha d\right)(0)
\right] + \ldots \nonumber \\
\eeqar
The ellipses on the right hand side denote gamma-structures other than
$\gamma_\varphi\gamma_5$ that do not contribute to $W^{\cal B}$.  Furthermore
in the above formula we did not write explicitly
the operator $<\bar{q} q> ( q \sigma^{\mu\nu}
g\tilde{G}^{\varrho\sigma} D^\alpha q)(0)$ due to the complicated
and lengthy structure of its coefficient.  In the end its contribution to the
final sum rule turns out to be small.

The relevant BPC arise from the correlation function of the local
operators present on the right-hand side of eq.~( \ref{etaope}) with
the twist-4 operator which defines $f^{(2)}$ at zero momentum.
In the particular case of
operators which are proportional to equations of motion of QCD, like
the operators $\bar q \gamma_5 \gamma^\varrho G_{\varrho\sigma}q$,
$\bar q \gamma_5 \sigma_{\varphi\lambda} D_\alpha q$ and $\bar q
\gamma_5 D_\alpha q$, the correlation functions can be evaluated by
means of exact low-energy theorems.  Using the functional integral
representation it is easy to obtain the following Ward identities, see
e.~g.~\cite{BK}:
\beqar{cterms}
&& i \int d^4y \langle 0| T O_\sigma(y)
\left(\bar q \gamma_5 D_\alpha q \right)(0)|0 \rangle
= 0
\nonumber \\
&& i \int d^4y \langle 0| T O_\sigma(y)
\left(\bar q \gamma_5 \sigma_{\varphi\lambda} D_\alpha q \right)(0) |0 \rangle
= m_0^2 <\bar q q> \frac{1}{12}
\left( g_{\sigma\lambda} g_{\alpha \phi} - g_{\sigma\phi} g_{\alpha \lambda}
\right)
\nonumber \\
&& i \int d^4y \langle 0| T O_\sigma(y)
\left(\bar q \gamma_5 \gamma^\varrho gG_{\varrho\alpha}q\right)(0)|0 \rangle
= \frac{4 \pi}{3} \alpha_S <\bar q q>^2 g_{\sigma \alpha}
\eeqar
To avoid misunderstanding we note that the above identities should be
understood in the following way. The large-distance contribution to
the lhs is equal to the large distance contribution to the rhs which
is just determined by a non-perturbative VEV of the corresponding operator.

The other remaining two-point correlation functions at zero-momentum
transfer cannot be evaluated exactly. Instead, they may be estimated
by considering additional two-point sum rules. Since ultimately one is
interested in their long-distance behaviour, we focus on the
contribution coming from the lowest possible intermediate state - the
massless, chiral, pseudoscalar meson. One can define the overlap of
the operators given in (\ref{etaope}) with a pion state as:
\beqar{pion}
&& \langle 0|\bar q \gamma_\mu \gamma_5 q |\pi(p)\rangle = i f_\pi p_\mu
\nonumber \\
&& \langle 0| \bar q g\tilde{G}_{\mu\nu} \gamma^\nu q |\pi(p)\rangle
= i f_\pi \delta_\pi^2 p_\mu
\nonumber \\
&& \langle 0| \bar q g\tilde{G}_{\mu\nu} D^\nu q |\pi(p)\rangle
= i f_\pi \tilde{\delta}^3_\pi p_\mu
\nonumber \\
&& \langle 0| \bar q \sigma_{\mu\varrho}g\tilde{G}^{\varrho\sigma} D_\sigma q
|\pi(p)
\rangle
= i f_\pi \bar{\delta}^3_\pi p_\mu
\eeqar
The constants $\delta_\pi$, $\tilde{\delta}_\pi$ and
$\bar{\delta}_\pi$ can be calculated by evaluating two-point
correlation functions in the sum rule framework at {\em non-zero}
momentum. We stress that to avoid double counting special care has to
be taken to subtract properly contributions of higher excited states
like the $A_1$-meson or the continuum.

Note that consideration of the contribution arising from pion exchange
alone is sufficient as long as we consider only the flavor non-singlet
combination $f^{(2)}(NS) = f^{(2)}(u) - f^{(2)}(d)$. Consideration of
the flavor singlet combination $f^{(2)}(S) = f^{(2)}(u) + f^{(2)}(d)$
naturally forces us to take into account contributions from $\eta$ and
$\eta^\prime$. In the chiral limit of massless quarks which we employ
in this paper the $\eta$ can be considered as massless. The axial
anomaly, however, will give rise to the $\eta^\prime$ mass
\cite{tHooft}. The pedestrian solution to this problem would be to
disregard the $\eta^\prime$ contribution as a short-distance one but
the real physics is certainly more complicated \cite{Ioffeano}. As our
calculation offers no insight into this complicated problem we have
decided simply to include the $\eta$ and the $\eta^\prime$ on the same
footing as the pion, but it should be kept in mind that due to the
axial anomaly the prediction for $f^{(2)}(S)$ is subject to an
uncertainty which is probably small, but presently unresolvable.

As far as $\delta_\pi$ is concerned an estimate due to
Novikov et.~al.~\cite{Novikov} can be found in the literature.
We essentially repeated
their calculation for correlators which determine $\tilde{\delta}_\pi$
and $\bar{\delta}_\pi$ with the result $\delta_\pi^2 = 0.21 \;\rm
GeV^2$, $\tilde{\delta}_\pi^3 = 0.033 \; \rm GeV^3$, $\bar{\delta}_\pi^3
= -0.1 \; \rm GeV^3$, $f_\pi = 133 \; \rm MeV$.
Hence, the
contribution of a massless chiral boson to the correlation
functions at zero momentum can be estimated as:
\beqar{pioncont}
&& i \int d^4y \langle 0| T O_\sigma(y)
\left(\bar q \gamma_\alpha\gamma_5 q \right)(0)|0 \rangle
= \frac{3}{4} f_\pi^2 \delta^2 g_{\alpha \sigma}
\nonumber \\
&& i \int d^4y \langle 0| T O_\sigma(y)
i \left(\bar q g\tilde{G}_{\alpha\mu}D^\mu q \right)(0)|0 \rangle
= - f_\pi^2 \delta_\pi^2 \tilde{\delta}_\pi^3 g_{\alpha \sigma}
\nonumber \\
&& i \int d^4y \langle 0| T O_\sigma(y)
\left(\bar q \sigma_{\alpha\varrho}g\tilde{G}^{\varrho\mu}D_\mu q \right)(0)|0
\rangle
e
= - f_\pi^2 \delta_\pi^2 \bar{\delta}_\pi^3 g_{\alpha \sigma}
\eeqar
The factor $\frac{3}{4}$ in the correlator of $O_\sigma$ with the axial
current is due to current conservation, see the discussion above. The
remaining correlation function $\Pi = i \int d^4y \langle 0| T
O_\sigma(y) O^\sigma(0)| 0 \rangle$ was evaluated with the help of an
additional sum rule, see \cite{BBK}, to be
$\Pi \sim 3 \cdot 10^{-3} \rm GeV^6$.
After Fourier transforming eq.~(\ref{etaope}) and inserting it
back into eq.~(\ref{ope}) we
finally obtain the expansion of $W^{\cal B}(p^2)$ in terms of both local and
bilocal power corrections:
\beqar{twist4ope2}
&& 4 W^{\cal B}(p^2)
= \nonumber \\
&&
A \frac{\alpha_S}{\pi^5} (-p^2)^4 \log(-p^2/\mu^2)
+ \left(B + B_{BL}\right) \frac{\alpha_S}{\pi} <\bar{q} q>^2 (-p^2)
\log(-p^2/\mu^2)
\nonumber \\
&&
+ C \frac{1}{\pi^4} <fg^3GGG> (-p^2) \log(-p^2/\mu^2)
\nonumber \\
&&
+ D \Pi \frac{1}{\pi^2} (-p^2) \log(-p^2/\mu^2)
+ F \frac{\alpha_S}{\pi} m_0^2<\bar{q} q>^2
\nonumber \\
&&
+ G \frac{<\bar{q} q>^2 <g^2G^2>}{-p^2}
+ \left(H + H_{BL}\right) \frac{<\bar{q} q>^2 m_0^4}{-p^2}
\nonumber \\
&&
+ I \pi \alpha_S <\bar{q} q>^2 f_\pi^2 \delta^2 \frac{1}{-p^2}
+
\left( J \tilde{\delta}_\pi^3 + K \bar{\delta}_\pi^3\right) <\bar{q} q>
f_\pi^2 \delta^2 \frac{1}{-p^2}
\eeqar
The numerical coefficients corresponding to the BPC can be read off
from table \ref{tabelle2}.

\begin{table}
\begin{center}
\begin{tabular}{||l|l|l|l|l|l|l||}
\hline \hline
  &  $B_{\rm BL}$  &  $D$ & $H_{\rm BL}$ & $I$ & $J$ & $K$ \\ \hline
d &  -20/ 27  & 5/36  &  -1/27  & $-64/9$ & $2^6/27$ &
5/9  \\ \hline
u &  -16/ 27  & 1/9   &   2/9   & 208/9 &  $2^5/27$ &
7/36 \\ \hline
\hline
\end{tabular}
\end{center}
\caption[]{\sf Numerical coefficients corresponding to BPC contributing to
the sum rule eq.~\ref{twist4ope2} .
The upper line gives the values for the twist-4 operator involving d quarks,
the lower line for the corresponding operator with u-quarks}
\label{tabelle2}
\end{table}

Before we proceed to extract the matrix element $f^{(2)}$ one
comment has to be made. An inspection of eq.~(\ref{twist4ope2})
reveals an important difference with respect to the previous
calculation of \cite{BBK}. The theoretical side of the present sum
rule given by eq.~(\ref{twist4ope2}) is manifestly free from effects of
mixing of the three-point correlation function (\ref{dreipunkt})
with two-point correlators \cite{BBK,BK}. In
other words the use of the quark-gluon current (\ref{qqqG}) resulted
in a much milder singularity structure of the three-point correlator than
in the case with the three-quark current. The mixing, which occurs
already at the tree-level i.e., before genuine radiative corrections
are considered, produces extra UV logarithms in Wilson coefficients and
makes it necessary to use a more complicated model of the spectral
representation \cite{Ioffe2}, although the resulting corrections turn
out to be much smaller than the overall uncertainties. In the present
case the representation (\ref{spec}) is sufficient to adequately
reproduce all  terms arising from the theoretical calculation.

To extract the matrix element of interest we employ the standard
strategy of QCD sum rules. First we multiply the sum rule by $m_N^2 -
p^2$ to eliminate the single-pole term in (\ref{spec}) and then apply
a Borel transformation to both sides, arriving at the following
expression
\beqar{sumtwist4}
&&-8 f^{(2)} \lambda_I \lambda_G m_N^6 e^{-m_N^2/M^2}
\nonumber \\
&&= \tilde A (5!\; E_6 M^{12} - 4!\; E_5 M^{10} m_N^2) +
        \tilde B (M^4 m_N^2 E_2 - 2 M^6 E_3) +
\nonumber \\
&&
        \tilde C (M^4 E_2 - M^2 m_N^2 E_1) + \tilde D m_N^2
\eeqar
where $\tilde A, \tilde B, \tilde C$ and $\tilde D$ are the coefficients in
front of $(-p^2)^4 \log(-p^2/\mu^2)$, $(-p^2) \log(-p^2/\mu^2)$,
$\log(-p^2/\mu^2)$ and $1/(-p^2)$ respectively.
$E_n(s_0,M^2)$ denotes as usual
\beq{eformel}
E_{n} = 1 - e^{\left(-s_0/ M^2 \right)} \sum_{k = 0}^{n-1}
\frac{1}{k!}\left(s_0 / M^2\right)^k  \, .
\eeq
The dependence on the overlap integrals $\lambda_I$ and $\lambda_G$
can be eliminated from eq.~(\ref{sumtwist4}) by dividing the
three-point sum rule by the two-point sum rule derived in \cite{BGMS}
\beqar{normierung}
&&2 (2 \pi)^4 m_N^2 \lambda_I\lambda_G  e^{-m_N^2/M^2}
= \frac{6}{5} \frac{\alpha_S}{\pi} M^8 E_4  \nonumber \\
&& + \frac{1}{2} <g^2GG> M^4E_2
   -\frac{4}{3}\frac{\alpha_s}{\pi}(2\pi)^4<\bar q q>^2 M^2E_1
   + \frac{2}{3}(2\pi)^4m_0^2<\bar q q>^2 \nonumber \\
\eeqar
The used set of condensates $<\bar{q}q> = (-0.257 \;{\rm
GeV})^3$, $<\alpha_s/\pi GG> = 0.012 \;{\rm GeV}^4$, $m_0^2 = <\bar{q}
g\sigma Gq>/ <\bar q q> = 0.65 \;{\rm GeV^2}$ and $<fg^3GGG> = 0.046
\;{\rm GeV^6}$ correspond to the standard ITEP values rescaled to the
normalization point $\mu^2_0 \sim m_N^2 \sim 1 \;{\rm GeV^2}$.  The
strong coupling constant at $1 \;{\rm GeV}$ is taken to be $\alpha_S = 0.37$
($\Lambda = 150)$ MeV.  The continuum threshold is chosen as
$s_0=(1.5\;{\rm GeV})^2$ roughly corresponding to the Roper resonance
position.

\begin{figure}
\centerline{\psfig{figure=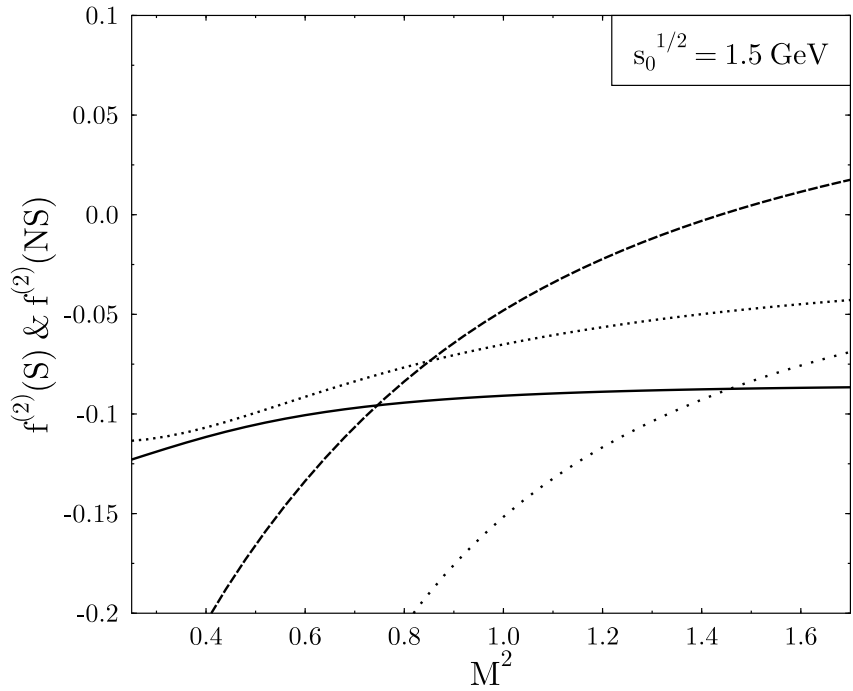,width=12cm}}
\caption[]{\sf Stability plot of the sum rule eq. (\ref{sumtwist4}).
The full line corresponds to $f^{(2)}(S)$, the
dotted line to $f^{(2)}(NS)$. For comparison the results of the analysis
in \cite{BBK} are also shown. The dashed line corresponds to
$f^{(2)}_{BBK}(S)$
the space-dotted line to $f^{(2)}_{BBK}(NS)$.}
\label{fig1}
\end{figure}
The quotient sum rule for the matrix element of the twist-4 operator has
been plotted in Fig.  \ref{fig1}. This figure shows the singlet (S)
and nonsinglet (NS) part of $f^{(2)}$.  $f^{(2)}(S) = f^{(2)}(u) +
f^{(2)}(d)$, $f^{(2)}(NS) = f^{(2)}(u) - f^{(2)}(d)$.  For comparison
the corresponding sum rules obtained from the analysis in
\cite{BBK} which employed Ioffe currents only are also shown.  The square of
the overlap integral $\lambda_I^2$ is determined from the additional
two-point sum rule
\beqar{ioffenorm}
&&2 (2 \pi)^4  \lambda_I^2  e^{-m_N^2/M^2}
=  M^6 E_3 \nonumber \\
&& + \frac{1}{4} <g^2GG> M^2E_1  + \frac{4}{3}(2\pi)^4<\bar q q>^2
\eeqar
which is just the standard sum rule considered by Ioffe.  Instead of
using a fixed value for $\lambda_I^2$ we divided the sum rule obtained
in \cite{BBK} by the sum rule (\ref{ioffenorm}).  The final values of
the matrix elements can be estimated from the figures by taking
$M^2\approx m_N^2 \approx 1$ GeV$^2 \approx \mu_0^2$, where $\mu_0^2$
represents the normalization point of the matrix element
(\ref{Osigma}). Numerically, their values are:
\beqar{numbers}
&&f^{(2)}(S) = - 0.09 \pm 0.02   \quad
f^{(2)}(NS) = - 0.07 \pm 0.02
\nonumber \\
&&f^{(2)}(proton) = -0.037 \pm 0.006  \nonumber \\
&&f^{(2)}(neutron) = -0.013 \pm 0.006
\eeqar
at $\mu_0^2 \approx 1$ GeV$^2$.  These values are to be compared with
those obtained from the sum rules given in \cite{BBK}:
\beqar{numbersBBK}
&&f^{(2)}_{\rm BBK}(S) = - 0.068 \pm 0.03 \quad
f^{(2)}_{\rm BBK}(NS)  = - 0.18 \pm 0.04
\nonumber \\
&& f^{(2)}_{\rm BBK}(proton) = - 0.049 \pm 0.01 \nonumber \\
&& f^{(2)}_{\rm BBK}(neutron)  = 0.01 \pm  0.01 \nonumber
\eeqar
All errors given are only due to the dependence on the Borel parameter
$M^2$. An additional error enters  due to the factorization of high
dimensional condensates. For condensates of dimension 8 the generally
accepted error is estimated to be of order $\sim 20 \%$.  There is
unfortunately very little experience with condensates of dimension 10
since such high dimensions occur seldom in calculations.
The good agreement of
our previous calculation \cite{SGMS} with the results of BBK \cite{BBK}
may be considered as a support for applicability of the factorization
procedure for dimension-10 condensates as well.  In our opinion an
estimated error of $\sim 50 \%$ is a very conservative guess.

Let us now discuss the importance of the various contributions
entering the expansion (\ref{twist4ope2}).
As in \cite{BBK,SGMS} the sum rule turns out
to be dominated by the operators of highest dimension i.e., those of
dimension 10. To be sure that this is a physical effect and not the
onset of a breakdown of OPE one should reliably estimate the next term
in the series, resulting in contributions of dimension 12, which is
clearly very difficult. Contrary to the evaluation of the twist-3
matrix element $d^{(2)}$ where BPC did not contribute, local and
bilocal corrections enter on equal footing in the present sum rule.
The numerically important BPC turn out to be those involving the
correlator with the axial current operator $\bar q \gamma_\alpha
\gamma_5 q$ and the correlator with the operator $\bar q \gamma_5
\sigma_{\varphi\lambda} D_\alpha q$ for which an exact low energy
theorem holds.  The contributions from $\tilde\delta^3_\pi$ and
$\bar\delta^3_\pi$ are numerically smaller and enter with opposite
signs so that they partially cancel each other.  Our final results
rely on the dimension-10 condensates $<\bar q q>^2 m_0^2$, $<GG>
<\bar q q>^2$ and the bilocal correction $<\bar q q>^2 i \int d^4y
\langle 0| T O_\sigma(y) \left(\bar q \gamma^\sigma\gamma_5 q
\right)(0) |0 \rangle$. This situation is different from
\cite{BBK} where the contribution of the dimension-8 condensate $<\bar q
q>^2 m_0^2$ plays the crucial role. It is by no means
trivial that a large number of different contributions merge together
to give a result similar to that of BBK.

As we have mentioned already the introduction of an explicit gluonic
component in the nucleon interpolating current resulted in a sum rule
which to the lowest order is free of extra UV logarithms due to
mixing, and therefore the additional uncertainties discussed in
\cite{BBK,BK,Ioffe2} do not influence the final estimate. In the previous
calculation of the twist-3 matrix element $d^{(2)}$ the mixing logarithm
arose in the dimension-8 contribution which was numerically negligible
in practice.  So, as expected beforehand, the consideration of
a non-perturbative gluonic component has lead to much less singular
behaviour of the correlators and therefore to more stable numerical
predictions.

In our previous calculation the values of the twist-3 matrix element
where found to be
$d^{(2)}(S) = - 0.068 \pm 0.03$ and
$d^{(2)}(NS) = 0.078 \pm 0.03$.
Using these numbers we find from eqn.~(\ref{Ocomp}, \ref{O3comp})
that both colour electric and colour magnetic fields in the rest system
of the nucleon contribute at the
same order of magnitude to the spin
\beqar{ebcomp}
\langle pS| gB^\sigma u^\dagger u |pS \rangle
&=& - (0.07 \pm 0.08) m_N^2 S^\sigma
\nonumber \\
\langle pS| gB^\sigma d^\dagger d |pS \rangle
&=& (0.188 \pm 0.08) m_N^2 S^\sigma
\nonumber \\
\langle pS|\left[(\bar u \vec{\gamma} \;u) \times g\vec{E}\;
\right]^\sigma|pS\rangle
&=& (0.09 \pm 0.08) m_N^2 S^\sigma
\nonumber \\
\langle pS| \left[(\bar d \vec{\gamma} \;d) \times g\vec{E}\;
\right]^\sigma|pS\rangle
&=& (0.21 \pm 0.08) m_N^2 S^\sigma
\; .
\eeqar
Obviously such a result shows that simple
phenomenological models motivated as analogy to QED are misleading.
In any such model one would expect the colour-magnetic term to dominate.

Finally we can analyze the
higher-twist contributions to the integral over $g_1^{p}(x)$ and
$g_1^{p - n}(x)$.
If we estimate $a^{(2)}$ from experiment,
$a^{(2)}(proton) = 0.022 \pm 0.002$ from EMC experiments \cite{EMC}
and $a^{(2)}(neutron) = 0.000 \pm 0.003$ from E142 \cite{SLAC}
we obtain for the Bjorken sum rule
\beq{bj}
\int_0^1 dx \; g_1^{p - n} (x,Q^2)
=
\frac{1}{6} a^{(0)}(NS) +
(0.003 \pm 0.008)\frac{\rm GeV^2}{Q^2} \; ,
\eeq
where the twist-2 matrix element is defined as
\beq{twist2}
2 S^\sigma a^{(0)}(f) =
\langle P S |\bar q_f(0) \gamma^\sigma \gamma^5  q_f(0) | P S\rangle
\; .
\eeq
In the case of the difference of proton and neutron structure function
the flavor index refers to the non-singlet combination $u - d$.
Using isospin symmetry this combination can be related to the
nucleon $\beta$-decay constant $g_A/g_V = a^{(0)}(NS)$ leading
to the celebrated Bjorken sum rule \cite{Bj}.
In case of the proton spin structure function we get
\beq{ej}
\int_0^1 dx \; g_1^{p} (x,Q^2)
=
\frac{1}{2} a^{(0)}(proton) -
(0.015 \pm 0.007)\frac{\rm GeV^2}{Q^2} \; .
\eeq
Here $a^{(0)}(proton)$ contains non-singlet and singlet combinations.
Using SU(3) symmetry it can be expressed by the $F$ and $D$
hyperon decay matrix elements.
To compare with the leading twist matrix elements we take the Bjorken
sum rule prediction  for $a^{(0)}(NS)$ and
for the Ellis-Jaffe leading twist element the measurement of
SMC \cite{SMC} taken at assumably asymptotic $Q^2$ .
\beqar{conl}
&&\int_0^1 dx \; g_1^{p - n} (x,Q^2 \to \infty) = \frac{1}{6}\frac{g_A}{g_V}
= 0.2095 \pm 0.0005 \nonumber \\
&&\int_0^1 dx \; g_1^{p} (x,Q^2 \sim 10 {\rm GeV^2}) =
0.136 \pm 0.011 \pm 0.011 \, .
\eeqar
Thus we can conclude that the present analysis suggests that the
higher-twist corrections to both sum rules are small for average $Q^2$
in the SMC and EMC range \cite{EMC,SMC}.

\bigskip

{\it Acknowledgements.} We would like to thank Vladimir Braun for many
usefull discussions and continous encouragement.  This work has been
supported by KBN grant 2~P302~143~06 nad by the German-Polish exchange
program X081.91.  A.S. thanks DFG (G.Hess Programm) and MPI f\"ur
Kernphysik in Heidelberg for support.

\vfill
\eject

\end{document}